# Secure Cross-Chain Provenance for Digital Forensics Collaboration

Asma Jodeiri Akbarfam[1], Gokila Dorai[1], Hoda Maleki[1]
[1]*School of Computer and Cyber Sciences, Augusta University* Augusta, USA
{ajodeiriakbarfam, gdorai, hmaleki}@augusta.edu

*Abstract*—In digital forensics and various sectors like medicine and supply chain, blockchains play a crucial role in providing a secure and tamper-resistant system that meticulously records every detail, ensuring accountability. However, collaboration among different agencies, each with its own blockchains, creates challenges due to diverse protocols and a lack of interoperability, hindering seamless information sharing. Cross-chain technology has been introduced to address these challenges. Current research about blockchains in digital forensics, tends to focus on individual agencies, lacking a comprehensive approach to collaboration and the essential aspect of cross-chain functionality. This emphasizes the necessity for a framework capable of effectively addressing challenges in securely sharing case information, implementing access controls, and capturing provenance data across interconnected blockchains. Our solution, ForensiCross, is the first cross-chain solution specifically designed for digital forensics and provenance. It includes BridgeChain and features a unique communication protocol for cross-chain and multi-chain solutions. ForensiCross offers meticulous provenance capture and extraction methods, mathematical analysis to ensure reliability, scalability considerations for a distributed intermediary in collaborative blockchain contexts, and robust security measures against potential vulnerabilities and attacks.Analysis and evaluation results indicate that ForensiCross is secure and, despite a slight increase in communication time, outperforms in node count efficiency and has secure provenance extraction. As an all-encompassing solution, ForensiCross aims to simplify collaborative investigations by ensuring data integrity and traceability.

## I. INTRODUCTION

Digital forensics plays a crucial role in investigations, allowing law enforcement agencies and organizations to extract, analyze, and preserve digital evidence for legal proceedings [1], [2]. However, ensuring the security of digital evidence, tracking the Chain of Custody (CoC), and maintaining the data provenance of the investigation, which involves tracing and authenticating the origin, custody, and history of any data artifact throughout the entire investigative process, remain primary challenges.

Recent advancements in blockchain offer a nuanced approach to addressing the perennial challenges of transparency, immutability, and security in evidence management. Unlike traditional methods, blockchain introduces a decentralized verification mechanism that inherently resists tampering, automates, and enforces the conditions for evidence access without compromising integrity. These make it an invaluable tool for not only law enforcement agencies but also sectors such as the Internet of Things (IoT), supply chain management, healthcare, and education, facilitating integration into organizational frameworks [3]–[5]. Blockchain adoption poses collaboration challenges for organizations due to the isolation created by independent private or public blockchains [6]. Cross-chain interoperability, introduced in 2014 by the Tendermint team, addresses this by enabling interoperability between blockchain ledgers [7]. However, achieving cross-chain transactions without a trusted third party necessitates secure solutions using centralized or decentralized trust mechanisms [8], [9]. Furthermore, structural differences in cross-chain processes require a unified approach to standardization to enhance functionality and scalability [10]. Managing cross-chain historical data, especially provenance, also presents significant challenges in overcoming data isolation to ensure accessibility for blockchain activities and analyses [10].

Cross-chain techniques are typically discussed in the context of asset transfer. However, some literature explores the use of relay chains as a prominent cross-chain technique designed to address various challenges faced by organizations collaborating with multiple blockchains. Existing literature, despite proposing effective relay chains like ARC [3], often overlooks factors such as the number of communication nodes and the scalability challenges of the relay, and fails to provide evaluations for the proposed methods. Solutions like Vassago [11], while effective for provenance queries across multiple blockchains, fail to address heterogeneity [12] and do not discuss provenance capture [13]. Therefore, to develop an effective solution, it is crucial to address relay chain design, communication, node count requirements, and evaluations that do not depend on the homogeneity of blockchains.

In the domain of digital forensics, collaboration among multiple law enforcement agencies or international cybercrime syndicates introduces additional complexities beyond the primary challenges associated with cross-chains. These include securely sharing case information, implementing effective access controls, capturing and analyzing data provenance across interconnected blockchains, extracting provenance details, advancing and synchronizing investigative stages, auditing access

This work was funded by NSF grant CCF-2131509

trails, and accommodating diverse analytical requirements. Existing approaches in digital forensics on blockchains [1], [14]–[18] have primarily focused on individual agencies. Research on inter-agency data sharing has typically involved a single blockchain for all nodes to join [19], indicating a lack of a comprehensive approach to collaboration, cross-chain functionality, and establishing provenance for tracking the chain of custody.

To address the mentioned challenges, we introduce ForensiCross, a secure framework for collaborative digital forensics across different blockchains. Our proposed architecture and algorithms aim to address the limitations of cross-chain methods for collaboration across multiple blockchains, such as provenance capture, heterogeneity, and relay design. We also consider the number of collaboration chains and incorporate domain-specific requirements for digital forensics, such as access control, provenance extraction and synchronization of investigative stages. We design Bridgechain as a distributed intermediary to address challenges in cross-chain communication, enabling seamless interactions between interconnected chains through a novel inter-blockchain communication protocol. The Bridgechain facilitates the exchange of data between heterogeneous private blockchains, each with its unique consensus mechanisms and security assumptions. ForensiCross employs nodes as validators across the involved blockchains, striking a balance between decentralization and oversight for forensic applications. In ForensiCross, when a case needs to be shared across blockchains, a formal agreement and meticulous logging of the process take place. This includes initiating a transaction on the source blockchain, activating a smart contract to generate a case smart contract, and processing the transaction through mutual nodes and the Bridgechain. After creating a case, ForensiCross ensures that only authorized query nodes can access the provenance by reaching an agreement among collaborating blockchains. It also implements an access control method specific to digital forensics, managing investigation stages and assigning roles with specific access privileges. In later phases, the system synchronizes the investigation process by forwarding stage transactions from one blockchain to the Bridgechain, requiring unanimous agreement from all involved blockchains for stage progression. The final phase allows each blockchain to handle its data retrieval and uploading by authorized users, ensuring comprehensive logging of digital forensics activities. A key aspect of ForensiCross is the extraction and verification of provenance information, achieved by constructing a novel Merkle tree for each blockchain involved in a case based on the investigation stages. This ensures a secure approach to provenance verification. ForensiCross addresses security concerns through a comprehensive analysis, mitigating threats such as mutual node compromise and provenance tampering. The evaluation emphasizes the significance of the Bridgechain as a decentralized intermediary, providing benefits such as fewer mutual nodes and secure collaboration. It also shows that scaling the number of mutual nodes with the increase in collaborating blockchains is a crucial aspect that has been overlooked in other literature. ForensiCross demonstrates its ability to adapt to the evolving landscape of digital evidence management and blockchain technology, ensuring robustness against threats like mutual node compromise and provenance tampering. The main contributions of this paper are as follows:

1) Developing an architecture specifically tailored for the collaboration and provenance of cross-chain solutions in digital forensics.
2) Introducing a novel communication protocol designed for the architecture of cross-chain and multi-chain environments.
3) Creating a provenance verification method aimed at efficient extraction of proof.
4) Conducting a novel analysis of the decentralized intermediary to adjust its node count in line with the growing collaboration among blockchains.
5) Evaluating the effectiveness of the proposed methods with mathematical analysis, decentralized intermediary lightweight system implementation, and security analysis.

This paper is organized as follows: Section II provides an overview of the background. Section III delves into related work, highlighting existing problems and challenges. Section IV discusses the significance of the methodology and proposed protocols. Section V includes a mathematical analysis of the framework, lightweight implementation, experiments, and security analysis. The paper concludes with Section VI.

## II. FUNDAMENTAL CONCEPTS

### A. Blockchain

A blockchain functions as a decentralized and distributed ledger, securely recording transactions across multiple network nodes [20]–[23]. It relies on mining and consensus algorithms to ensure network security [24]. Blockchain's core functionality fosters trust, transparency, and security in digital transactions. Immutability stands as a cornerstone of blockchain technology, guaranteeing data integrity and resistance to tampering. This is achieved through two essential components: the Merkle root and the hash of the previous block [25]. Additionally, blockchain cryptographically links blocks, such that any alteration to a previous block renders subsequent ones invalid [26], significantly enhancing the system's integrity.

There are two main categories of blockchains: public and private. Public blockchains, exemplified by cryptocurrencies like Bitcoin and Ethereum, are distinguished by their open accessibility to anyone. Participants can freely join the network, validate transactions, and contribute to maintaining the distributed ledger. In contrast, private blockchains restrict access to authorized participants, often within a specific organization. These private networks are designed to enhance privacy and provide control, particularly in enterprise settings [27].

### B. Cross-Chain

The design of distinct blockchain systems is significantly influenced by the diverse requirements of various applications,

introducing challenges to their interoperability. This complexity leads to the creation of isolated data segments, further complicating the process of connecting individual blockchain systems [12]. An illustrative example of this scenario involves institutions 1 and 2 deploying blockchain A and blockchain B separately. Users affiliated with Institution 1 aim to interact with Institution 2's blockchain B, resulting in a cross-chain interaction model. The primary objective of this interaction, as detailed by Ou *et al.* [6], is to ensure both authenticity and credibility. Cross-chain functionality emerges as a solution to overcome these challenges, enabling the seamless transfer of data and assets across distinct blockchain networks. This facilitates the connection of various blockchains, ultimately resolving the complexities associated with isolated data segments, a challenge often referred to as value islands [28]. Current cross-chain systems primarily rely on notary schemes [29], hash-locking techniques [30], sidechains [31], [32], or relay chains [33], [34] where in some literature side and relay chains are used interchangeably. Notary schemes use a third-party intermediary to facilitate transactions between blockchain chains lacking trust, with examples like the InterLedger protocol (ILP) [30]. Hash time-locked contracts (HTLCs) streamline asset exchanges across blockchains, ensuring atomic swaps without trusted intermediaries [35], [36]. Sidechains run parallel to the main chain, enhancing performance and extending capabilities [37], [38]. Relays establish links between different chains, supporting various use cases like asset portability and atomic swaps [39], [40].

*C. Digital Forensics*

Digital forensics follows a structured five-stage methodology: identification, preservation, collection, analysis, and reporting. Investigators first identify evidence sources and relevant individuals. They then preserve Electronically Stored Information (ESI) to prevent data alteration. Next, digital data is collected, and exact duplicates are created for detailed analysis. Finally, findings are compiled into a comprehensive report. This process ensures evidence integrity and legal admissibility, adhering to standards set by organizations such as NIST [41].

III. RELATED WORK

*A. Blockchain for Data Provenance*

Blockchain technology has been extensively investigated for its application in recording data provenance across diverse domains, encompassing general data protection regulation (GDPR) data collections, IoT, supply chain management, machine learning, cloud computing, scientific workflows, legal scenarios, and digital forensics [42]–[48]. Noteworthy systems such as LineageChain and BlockCloud focus on detecting data modification attempts and implementing efficient query techniques and consensus protocols, while ProvHL emphasizes access control management and user consent mechanisms [49]–[51]. The concept of provenance holds particular significance in scientific workflows, with various works like Block-Flow, SciLedger, SmartProvenance, DataProv, Nizamuddin *et al.*, SciBlock, Bloxberg, and SciChain introducing specialized approaches incorporating event listeners, voting systems, decentralized databases, timestamp-based invalidation, and unique provenance models [48], [52]–[58]. The IoT expansion has been remarkable across various domains in recent years. Notably, blockchain-based provenance mechanisms have been incorporated in the IoT domain to ensure integrity and verifiability through transaction records within the blockchain network [59]. For instance, Pahl *et al.* [60] integrated IoT edge orchestrations with blockchain-based provenance, addressing trust concerns by recording origin and actions in the blockchain network. Javaid *et al.* presented BlockPro [61], a secure IoT framework utilizing blockchain for data provenance and integrity, and Ali *et al.* [62] proposed a secure provenance framework for cloud-centric IoT, incorporating blockchain for identifying data origin. Provenance records in digital forensics are crucial for preserving evidence integrity, and the IoTFC framework addresses IoT-specific forensic challenges. However, it has limitations such as neglecting access control, lacking clear component communication, and not effectively evaluating the applicability and data extraction [1]. Several proposed solutions in the field of digital forensics aim to enhance investigative processes. Akbarfam *et al.* [5] introduced Forensiblock, a private blockchain incorporating an access control method specifically designed for digital forensics. This system focuses on tracking the provenance of investigations and extracting all relevant information. Borse *et al.* [63] presented a hybrid blockchain solution with a primary emphasis on CoC management. Additionally, Ahmed *et al.* [17] suggested a Hyperledger-based private blockchain and IPFS system designed for tracking media files as evidence.

*B. Cross-Chain Methods*

In the study by Wu *et al.* [10], the cross-chain workflow model revolves around a relay chain that acts as a decentralized and trustworthy intermediary linking notaries and side-chains. This model effectively facilitates interactions among various blockchains, offering technical interoperability for cross-chain transactions. The use of workflows, abstractions, and specifications ensures replicability and well-defined cross-chain processes with three distinct workflow types: consensus, execution, and query workflows. Zhang *et al.* [3] present ARC, a relay chain system tailored for consortium blockchain environments. ARC operates on Hyperledger Fabric and employs an asynchronous consensus protocol for enhanced resilience and scalability. The relay chain serves as a central transaction hub, connecting application chains and streamlining interactions for seamless cross-chain transactions. Ding *et al.* [64] propose an extensible cross-chain access control and identity authentication scheme for consortium blockchain systems. The scheme ensures authenticated cross-chain operations and scalability improvements by leveraging a relay chain-based framework. Mutual authentication, access control, and identity authentication are incorporated during chain registration and data circulation processes. Chang *et al.* [65] introduce SynergyChain, a blockchain-driven framework for the secure sharing of patient electronic medical records (EMRs) across diverse blockchain networks. The architecture includes three tiers, addressing privacy challenges through standardized data submission, hierarchical access control, and additional layers of security, such as block header synchronization and a data validator sub-module. Han *et al.* [11] present Vassago, an innovative multi-chain system designed to improve the efficiency and credibility of cross-chain provenance queries. The architecture includes layers for provenance tracking, smart contracts, two-layer storage, and network interactions, focusing on principles to prevent tampering with cross-chain transactions and ensure reliable relevance among nodes. Vassago focuses on cross-chain transaction dependencies, validating authenticity through a shared blockchain, and parallelizing query processes.

*C. Problems and Challenges*

The challenges of establishing provenance within a single blockchain are detailed in Section III-A, and these challenges intensify in cross-chain scenarios. Existing frameworks, though beneficial, lack comprehensive solutions for provenance and cross-chain requirements, such as robust security measures, access controls, and streamlined provenance capture and extraction, particularly in cross-chain communication. Despite efforts to address some issues, capturing retail provenance and ensuring seamless cross-blockchain communication remain unresolved. Additionally, there is no dedicated framework for digital forensics, complicating the alignment with trust assumptions and requirements. Digital forensics requires secure case information sharing, robust access controls, meticulous data provenance capture and analysis across interconnected blockchains, extraction of detailed provenance information, synchronization of investigative stages, auditing of access trails, and adaptation to diverse analytical and organizational needs.

## IV. FORENSICROSS FRAMEWORK

The ForensiCross framework orchestrates a symbiotic collaboration among diverse entities, as illustrated in Figure 1 This setup resembles the process when blockchain A initiates a collaboration with blockchain B, and it can extend to include multiple collaborating blockchains. The essential entities and their corresponding notations are detailed as follow:
**Bridgechain:** Functioning as a blockchain intermediary, the Bridgechain plays a critical role in routing and facilitating communication between interconnected chains. It actively maintains the integrity and security of digital evidence. In the context of digital forensics, it refers to the private blockchain of the organization that is trusted by all others and has Proof of Authority (POA) mining.
**Blockchains:** Each organization involved in a digital forensics investigation operates on its private blockchain, ensuring the security and integrity of digital evidence within the framework. Each blockchain can function as either a source or destination for data.
**Users:** Authenticated users within each blockchain have the ability to send transactions. Each user has a set of Public and Private key and this category includes specific users empowered to conduct comprehensive queries across the entire digital forensics framework, known as Query users.
**Trusted Nodes:** Nodes positioned within each entity bear the responsibility of processing digital evidence and upholding the integrity of the digital forensics system. These nodes also function as miners of the blockchains, utilizing POA mining.
**Mutual Nodes:** Selected trusted nodes from each blockchain also serve as trusted Nodes of the Bridgechain. They act as authenticated entities within the digital forensics context, capable of mining on both the blockchain and the Bridgechain. Their selection is a result of agreement between the organization responsible for the Bridgechain and the organization responsible for the blockchain.

*A. Motivation and Example*

To illustrate the goals of ForensiCross, consider a complex murder case involving multiple jurisdictions. In this scenario, collaboration among police agencies is crucial as they gather evidence and conduct analyses. To facilitate this collaboration without major modifications to their existing private blockchain systems, agencies require a solution that can track COC, synchronize investigation stages, validate trust assumptions, enforce access control, and securely exchange information. The ForensiCross framework is tailored to meet these needs. The role of the Bridgechain in ForensiCross, compared to other relay mechanisms, is pivotal. Acting as a trusted intermediary, the Bridgechain ensures seamless communication, data integrity, and collaboration among the involved agencies. It can be implemented as a third-party trusted agency specifically chosen for the investigation, or it could be a designated entity within one of the participating agencies.

*B. Inter-Blockchain Communication*

A crucial aspect of cross-chain communication involves facilitating interactions between distinct blockchains, allowing organizations to collaborate securely with an immutable method. As each blockchain operates with unique protocols, ensuring effective communication and trust among these diverse networks necessitates the use of a Bridgechain as an intermediary. Acting as a conduit, the Bridgechain utilizes mutual nodes proficient in translating the distinct language of each blockchain, enabling seamless data transfer and synchronization. To facilitate this, communication smart contracts are employed on each blockchain and the Bridgechain.

When a transaction is initiated on the source blockchain, a series of orchestrated steps ensures its successful propagation to the destination blockchain, as depicted in Algorithm 1. To provide a more detailed understanding, the process is elaborated below:

1) **Transaction Initiation and Identification:** A user on the source blockchain initiates a transaction, embedding within it, Public Key, the identification of the destination blockchain, and signing it.
2) **Smart Contract Intervention:** The communication smart contract on the source blockchain, after validat-

ing the user's signature and the transaction, assigns a unique identification to the transaction and facilitates its submission.

3) **Mutual Node Translation:** The set of mutual nodes belonging to the source blockchain and Bridgechain detects the transaction. Each node undertakes the intricate task of translating the transaction into the Bridgechain's standardized format, subsequently submitting it to the Bridgechain.

4) **Consensus Verification and Validation:** The communication smart contract on the Bridgechain actively monitors incoming transactions, cross-referencing their content and tallying their numbers. It evaluates the accuracy of these translations and keeps track of transaction counts using a verification function and a counting function. Once it confirms that more than half of the mutual nodes have submitted the same translated transactions, the communication smart contract validates the transaction and promptly records it on the blockchain.

5) **Destination Blockchain Integration:** The trusted nodes between the Bridgechain and the destination blockchain receive the validated transaction by checking the Bridgechain, further translating it to the destination blockchain's specifications, and seamlessly incorporating it into the chain.

This process ensures the interoperability of diverse blockchains with a focus on reliability and integrity. The same approach is consistently applied in subsequent stages whenever reference is made to a blockchain communicating with external entities.

---

**Algorithm 1:** Inter-Blockchain Communication

**Data:** Initial Transaction on the source blockchain (SB)
**Result:** Final transaction on the destination blockchain (DB)

$T \leftarrow$ INITIATEUSERTRANSACTION($PK, SB, DB, Sig$);
SB−CSC.VALIDATEUSERSIGNATURE($T$);
($T, ID, SP$) $\leftarrow$ SB_CSC.INITIATETRANSACTION($T$);
$T_{new} \leftarrow$ CREATENEWTRANSACTION($T$);
MUTUALNODESTRANSLATE($T_{new}$);
BC_CSC.MONITORTRANSACTIONS();
BC_CSC.VERIFYTRANSACTIONS();
MUTUALNODESTRANSLATE($T_{new}$);
DB−CSC.MONITORTRANSACTIONS();
DB−CSC.VERIFYTRANSACTIONS();

---

### C. System Phases

ForensiCross operates through distinct phases, each designed to meet the requirements of digital forensics collaboration.

*1) Cross-Chain Digital Evidence Case Creation:* In this phase, when a case needs to be shared across blockchains, a formal agreement and initiation is essential, with the entire process meticulously logged. The following steps outline the process of creating a shared case:

1) Case Create Request: A user initiates the process by submitting a transaction to the source blockchain which contains information such as case number, destination blockchain, and signature.

2) Smart Contract Activation: Upon receiving the transaction, the communication smart contract activates and generates a case smart contract. This contract preserves crucial details, including the case number, all destination blockchains involved, and the creator's public key. The case smart contract is useful for tracking every case locally in each blockchain.

3) Communication: The communication smart contract issues a transaction on the blockchain, observed by all trusted nodes between the source blockchain and the Bridgechain. These nodes convert the transaction into the Bridgechain format and post it on the Bridgechain.

4) Bridgechain Processing: The communication smart contract on the Bridgechain processes this information. Upon transaction approval, a new digital case and corresponding smart contract are established. The source and destination blockchains are duly recorded, and communication transactions for the destination blockchains are initiated.

5) Mutual Node Translation: The mutual nodes of the destination blockchain since they translate the transaction into their native format before transmitting it to the destination blockchain. The communication smart contract on the destination blockchain oversees incoming transactions, facilitating the creation of a new smart contract for the shared case while referencing the source blockchain.

*2) Access Control Method:* Effective data retrieval and processing in various analyses require comprehensive process logging and appropriate user access levels. In the realm of digital forensic cases, a Role-Based Access Control with Staged Authorization (RBAC-SA) model, as detailed in [5], is implemented upon case creation. Managed by the source blockchain, this model delineates investigation stages and assigns roles with specific access privileges. Access control is enforced through a dedicated transaction dispatched from the originating blockchain, specifying roles and their associated access rights at each stage.

*3) Query Node Assignment:* In digital forensics, not everyone can query the provenance. In this scenario, the different blockchains collaborating on a case must reach an agreement on the query nodes. While the query of a local blockchain depends on the organization's decisions, for cross-chain queries, each blockchain sends a transaction to the Bridgechain. This transaction contains the public keys of the query nodes of that blockchain and the case number. The Bridgechain then adds the public keys as query nodes to the case.

*4) Investigation Stage Progress:* To synchronize the investigation process, users of one of the blockchains involved in the investigation send a *Stage* transaction to their blockchain, which is then forwarded to the Bridgechain. The Bridgechain adds this information to the case smart contract and forwards it to destination blockchains. Each destination blockchain votes on the case stage progress and forwards the information to the

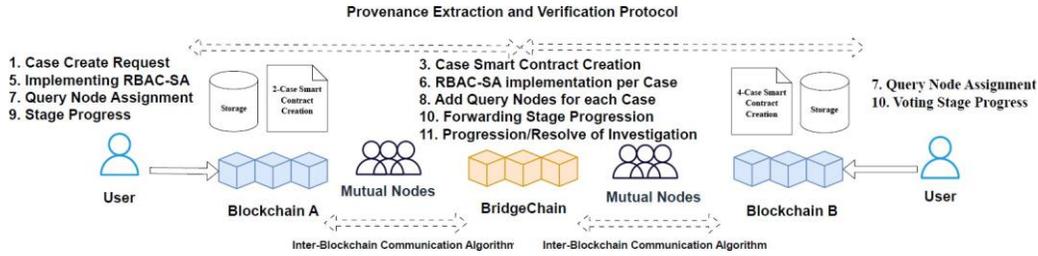

Fig. 1: ForensiCross Framework

Bridgechain. The Bridgechain's communication smart contract checks if all blockchains agree; if they do, confirmation is sent to the blockchain for stage progression. If not, a transaction is sent detailing the issue and requesting resolution. Due to the nature of digital forensics, the resolution is carried out offline, and once resolved, the request is repeated on the blockchain, requiring unanimous agreement from all votes.

*5) Data Retrieval and Uploading Procedures:* While cases are shared, each blockchain manages its own data retrieval and uploading procedures by authorized users. It is essential that all requests are processed by the blockchain for logging purposes, ensuring a comprehensive record of digital forensics activities.

### D. Provenance Extraction and Verification Protocol

The integration of blockchains plays a pivotal role in capturing detailed logs related to case initiation, access, edits, and collaborations within the ForensiCross framework. However, managing the substantial volume of logs and addressing privacy concerns in digital forensics necessitates the secure transmission of these logs. To facilitate the efficient extraction of provenance information from any blockchain without the need to query the entire blockchain, transactions for each case are stored in off-chain storage. Despite the establishment of trusted nodes between the blockchain and the Bridgechain to ensure reliability, there is a potential compromise of trust when utilizing off-chain storage for provenance information extraction. To mitigate this risk, a hash of the transaction is generated by mutual nodes and transmitted to the Bridgechain whenever a case request is submitted within the source blockchain. The Bridgechain then maintains a hash record of all transactions at various stages across the involved blockchains. As illustrated in Figure 2, the Bridgechain constructs a Merkle tree for each blockchain involved in the case, with the leaves representing the investigation stages. Upon receiving a provenance request from one of the query nodes in the blockchains, the request is forwarded to each destination blockchain after verifying the authenticity of the query node. The respective blockchains retrieve the provenance information from their off-chain storage, encrypt the information using the public key of the query node, and transmit it through a secure channel. Therefore, while the request for provenance is initiated within the blockchain, the data is sent off-chain. Since the Bridgechain possesses the hash record of every stage of the investigation, it appends this information and transmits the consolidated data to the requesting blockchain. Subsequently, the information is relayed to the query node, and its authenticity can be verified. Upon receiving the Merkle root, the provenance records for each blockchain can be verified using this method, and any tampering can be identified, specifying which stages were affected.

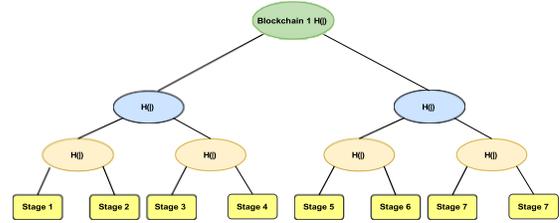

Fig. 2: Provenance Per Involved Blockchain

## V. EVALUATION

We have conducted a comprehensive evaluation of the framework, focusing on the utilization of a Bridgechain. A comparison is drawn between scenarios with direct connections between blockchains and those utilizing a Bridgechain, supported by mathematical formulas and constraints to justify the necessity of the latter. The evaluation encompasses aspects such as the implementation of a communication protocol, security analysis, and the results of the evaluation, emphasizing the framework's efficiency and security measures against threats like mutual node compromise and provenance tampering.

### A. Bridgechain Structure as an Intermediary

The introduction of a Bridgechain as a decentralized intermediary between blockchains is deemed essential for enhancing trust and maintaining provenance. However, contemplating scenarios where trusted nodes establish direct connections between each pair of blockchains shown inwithout relying on a Bridgechain requires closer scrutiny. These mutual nodes, potentially are in both blockchains, facilitate direct transaction translation. As illustrated in Figure 3a using this method can lower the number of communications as opposed to Bridgechain, however as shown in Figure 3b, the number of mutual nodes experiences a significant increase with the addition of more interconnected blockchains. To justify the number of mutual nodes needed in both scenarios and the size of the Bridgechain a comprehensive comparison of these two design approaches is needed, investigating the use of Bridgechain for reasons that extend beyond the initial considerations. The ensuing analysis aims to shed light on the constraints of both method's advantages offered by a Bridgechain and further justifies its role in fostering secure and efficient cross-chain communication. Thereby, let $k$ be the number of blockchains, $m$ the total number of nodes in the Bridgechain, $n$ the number of nodes in each blockchain, $i$ the $i$-th blockchain, $n_i$ the number of mutual nodes for the $i$-th blockchain, and $b_i$ the number of mutual nodes for the Bridgechain. In establishing a robust and reliable communication protocol, we implement specific constraints within the ForensiCross framework. A pivotal rule dictates that the minimum number of translated transactions ($n_i$) must exceed 2. This stipulation is carefully designed to eliminate the risk of a single point of failure in communication. Additionally, our intentional choice of an odd number for $n_i$ underscores our commitment to the protocol's integrity. This strategic decision significantly enhances the reliability of ForensiCross, fortifying it against potential failures and upholding the integrity of cross-blockchain communication.

Moreover, we introduce an additional rule to reinforce the robustness of the Inter-Blockchain communication protocol. The rationale behind this rule is to prevent a scenario where mutual nodes, being a minority, could potentially influence the approval of transactions. By ensuring that mutual nodes neither dominate the process nor operate as a self-approving entity, this constraint adds an extra layer of security to the ForensiCross framework.

$$2 < n_i < \frac{m}{2} \quad (1)$$

$$2 < n_i < \frac{n}{2} \quad (2)$$

Additionally, to maintain exclusivity, we introduce the exclusion constraint for mutual nodes:

$$n_{i+1} \cap n_i = \varnothing \quad (3)$$

This constraint guarantees that in the $(i+1)$-th blockchain, there are $n_{i+1}$ mutual nodes that do not overlap with the previous $n_i$ mutual nodes. For further analysis, we inspect the two cases separately: Mutual nodes between blockchains and Mutual nodes with Bridgechain.

**Mutual nodes between blockchains:** To enable communication among multiple blockchains, a complete graph formation is required, meaning each blockchain must be connected to every other blockchain. The formula for the number of edges in a complete graph, as well as the minimum number of mutual nodes (considered to be 3 for simplicity), is given by:

$$ni = \frac{k \cdot (k-1) \cdot 3}{2} \quad (4)$$

**Mutual nodes with Bridgechain:** The mutual nodes establish trust between blockchains by being employed between each blockchain and Bridgechain. For this scenario, adding another constraint:

$$2 < |b_i| \leq \min\left(\frac{1}{2}|n|, \frac{1}{2}|m|\right) \quad (5)$$

Here, we ensure that the absolute value of $b_i$ is at least 3 and does not exceed the minimum of half the number of nodes in each blockchain or half the total number of nodes in the Bridgechain.

The requirement for least mutual node count decisions is driven by the communication protocol's needs. In this protocol, more than half of the mutual nodes must agree on the same transaction for it to be considered valid. Having at least an odd number of mutual nodes helps prevent ties and ensures a clear majority decision. Additionally, this assumption provides some level of resilience against malicious behavior or incorrect information from a single node. The upper limit on mutual nodes is set to avoid inefficiencies or conflicts in consensus within the blockchains, as mutual nodes are also part of the individual blockchains and run smart contracts.

Furthermore, based on the constraints, the conclusion for the total number of nodes and mutual nodes in Bridgechain is:

$$m \geq 6k + 1 \quad (6)$$

$$bi \geq 3k \quad (7)$$

After computing the total number of mutual nodes for both cases we will have :

$$3k \leq \frac{k \cdot (k-1) \cdot 3}{2} \quad (8)$$

$$3 \leq k \quad (9)$$

Therefore, it can be concluded that, when multiple blockchains collaborate, the Bridgechain offers significant benefits in terms of the number of mutual nodes. However, both approaches for ensuring reliability and other factors require essential limitations. Additionally, we emphasize the importance of a decentralized intermediary like the Bridgechain to adjust its node count in line with the growing collaboration among blockchains.

*B. Communication Protocol*

To assess the Communication Protocol's effectiveness and case creation between multiple blockchains, we developed a lightweight system prototype in Python. This prototype features private blockchains built from scratch, each with distinct nodes running on various ports, and utilizes Flask for networking. The choice of private blockchains for digital forensics was motivated by security concerns, ensuring the isolation and security of the blockchains from external interference. Our prototype demonstrates that these blockchains can collaborate seamlessly without requiring any changes to their structure or underlying algorithms, highlighting the flexibility and compatibility of the Communication Protocol. Testing was conducted on a device equipped with an 11th Gen Intel(R) Core(TM) i7-1165G7 processor running at 2.80GHz and 16.0

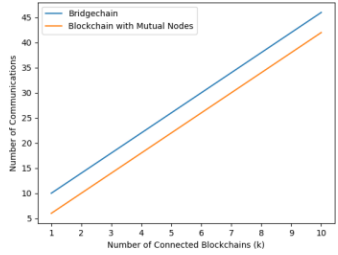
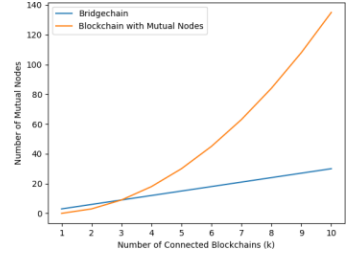

(a) Number of Communications   (b) Minimum Number of Mutual Nodes

Fig. 3: Comparison of Communication Efficiency and Mutual Node Distribution

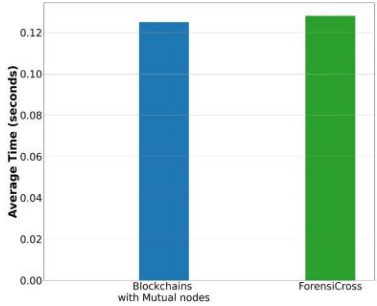

Fig. 4: Communication Duration

GB of RAM. In our experiments, we simulated various activities within the system by manipulating the number of blocks and cases. A comprehensive list of transactions was generated to mimic real-world scenarios, with some cases exclusively residing on individual blockchains and others shared between multiple blockchains. These transactions were transmitted to the network for processing, allowing us to analyze the system's performance and behavior. To accurately simulate communication, we eliminated the block creation time of the source blockchain. Instead, we measured the time from when the mutual nodes received the transaction until the smart contract on the destination blockchains accepted it. This ensured that the communication time accurately reflected the process of transmitting and validating transactions between blockchains. Additionally, we dedicated the same amount of time to the transaction time each time a transaction was needed, ensuring consistency in the communication process across different scenarios and transactions.

Our experiments compared scenarios with blockchains connected without a Bridgechain and only with mutual nodes. The results, as shown in Figure 4, indicate that while case creation took longer with ForensiCross than with blockchains with mutual nodes due to the increased count of communication, as illustrated in Figure 3a, the difference in time was not significantly larger.

### C. Security Analysis

We address two key security risks in ForensiCross: node compromise and provenance tampering. These threats undermine cross-blockchain communication integrity and digital forensics data preservation. We detail how ForensiCross mitigates these risks to ensure transaction and provenance record security.

*1) Mutual Node Compromise:* The compromise of mutual nodes poses a potential threat, particularly concerning their role in translating transactions across diverse blockchains. This compromise introduces intentional inaccuracies in transaction translations or alterations to their destinations.

**Mitigation:** ForensiCross employs a communication smart contract that actively monitors, verifies, and validates translated transactions by mutual nodes. It incorporates a verification mechanism as follows for the donating $T$ as translated transactions and $n$ as the number of total translated transactions:

$$\text{Verify} = \begin{cases} 1 & \text{if } T_i = T_{i+1} \text{ for } i = 1, 2, \ldots, \frac{n}{2} \\ 0 & \text{otherwise} \end{cases}$$

In this mechanism, more than half of the mutual nodes should have malicious translations, thereby being corrupted nodes. Furthermore, operating within a private blockchain environment with the POA consensus mechanism further fortifies the framework, substantially reducing the likelihood of mutual node compromise.

*2) Provenance Tampering:* The compromise of the integrity of the provenance records storage occurs, resulting in the tampering of records for a specific case.

**Mitigation Strategy:** ForensiCross employs an approach to mitigate provenance tampering. Query nodes play a pivotal role in ensuring the integrity of provenance records by reconstructing the Merkle root from stored records and comparing it with the Merkle root obtained from the Bridgechain. This process enables the query node to pinpoint the specific stage and blockchain implicated in the investigation, thereby validating the integrity of provenance records. For each stage, the hash and The Merkle root is computed :

$\text{Stage}_i = H\left(H(\text{Transaction}_1, \text{Transaction}_2, \ldots, \text{Transaction}_n)\right)$

$H\left(H(\text{Stage}_1, \text{Stage}_2), H(\text{Stage}_3, \text{Stage}_4), \ldots, H(\text{Stage}_7, \text{Stage}_7)\right)$

This computed Merkle root is then compared with the Merkle root obtained from the Bridgechain. If any discrepancy arises, indicating a potential compromise, a systematic verification is performed, scrutinizing each stage's hash individually to precisely identify the compromised stages and in the event of verification failure, further granular comparison can be conducted stage by stage:

$$\text{CompareMerkleRoot}(\text{MerkleRoot}_{\text{Bridgechain}}, \text{MerkleRoot}_{\text{Local}})$$

$$\text{CompareStage}(\text{Stage}_i, \text{BridgechainStage}_i)$$

This process ensures the detection and identification of compromised stages by scrutinizing both the computed Merkle root and individual stage hashes in comparison with those from the Bridgechain.

*D. Evaluation Overview*

As the number of collaborating blockchains increases, Forensicross manages to handle the escalation in communications efficiently. Despite the increase, the marginal time difference compared to an alternative approach without the Bridgechain is overshadowed by the greater efficiency in mutual node count and secure provenance extraction. The framework is resilient against mutual node compromise and provenance tampering. It also demonstrates that scaling the number of mutual nodes as the number of collaborating blockchains increases is necessary.

## VI. CONCLUSION

In digital forensics, precise provenance records are crucial for evidence credibility and integrity, fostering agency collaboration. However, current research focuses on individual agencies. ForensiCross addresses this by enhancing digital forensics through collaboration and cross-chain functionality. It rectifies inefficiencies, establishes interoperability, addresses security challenges, amplifies evidence traceability, enables cross-blockchain communication, and facilitates secure provenance extraction. ForensiCross's architecture includes Bridgechain, blockchains, users, trusted nodes, and mutual nodes, operating through phases like case creation and access control. It proposes a novel protocol for inter-blockchain communication and a method for provenance record extraction and verification.